\documentclass[runningheads]{llncs}
\usepackage{graphicx}
\usepackage{hyperref}
\begin{document}
\newtheorem{defi}[theorem]{Definition}
\title{Property Graph Exchange Format}
%
%
\author{Hirokazu Chiba\inst{1} \and Ryota Yamanaka\inst{2} \and Shota Matsumoto\inst{3}}
\authorrunning{H. Chiba et al.}
%
\institute{
Database Center for Life Science, Chiba 277-0871, Japan\\
\email{chiba@dbcls.rois.ac.jp}
\and
Oracle Corporation, Bangkok 10500, Thailand\\
\email{ryota.yamanaka@oracle.com}
\and
Lifematics Inc., Tokyo 101-0041, Japan\\
\email{shota.matsumoto@lifematics.co.jp}
}
\maketitle              
\begin{abstract}
Recently, a variety of database implementations adopting the property graph model have emerged. However, interoperable management of graph data on these implementations is challenging due to the differences in data models and formats.
Here, we redefine the property graph model incorporating the differences in the existing models and propose interoperable serialization formats for property graphs. The model is independent of specific implementations and provides a basis of interoperable management of property graph data. The proposed serialization is not only general but also intuitive, thus it is useful for creating and maintaining graph data. To demonstrate the practical use of our model and serialization, we implemented converters from our serialization into existing formats, which can then be loaded into various graph databases. 
This work provides a basis of an interoperable platform for creating, exchanging, and utilizing property graph data.
\end{abstract}

\section{Introduction}
Increasing amounts of scientific and social data are described and analyzed in the form of graphs. In the context of graph analysis, the property graph model~\cite{angles} is becoming popular; various graph database engines, including Neo4j~\cite{neo4j}, Oracle Labs PGX~\cite{pgx}, and Amazon Neptune~\cite{neptune}, adopt this model. These graph database engines support powerful algorithms for traversing or analyzing graphs. 
In contrast to the standardized RDF, however, property graphs lack a standardized data model.
Here, we considered the general requirements for representing property graphs and designed two serialization formats as flat text and JSON. These formats can be converted into specific formats for each of the databases mentioned above. The serialization formats independent of certain database implementations will increase the interoperability of graph databases and will make it easier for users to import accumulated graph data.

\section{Model Definition}
Here, we define the property graph model independent of specific graph database implementations. For the purpose of interoperability, we incorporate differences in property graph models, taking into consideration multiple labels or property values for nodes and edges, as well as mixed graphs with both of directed and undirected edges. The property graph model we redefine here requires the following characteristics:

\begin{itemize}
    \item A property graph contains nodes and edges.
    \item Each of the nodes and edges can have zero or more labels.
    \item Each of the nodes and edges can have properties (key-value pairs).
    \item Each property can have multiple values.
    \item Each edge can be directed or undirected.
\end{itemize}
More formally, we define the property graph model as follows.

\begin{defi}[Property Graph Model]
\leavevmode \vspace{1mm} \\
A \emph{Property Graph} is a tuple
$PG = \langle N, E_d, E_u, S, V, P, e, l_n, l_e, p_n, p_e\rangle$, where:
\begin{enumerate}
    \item $N$ is a set of nodes.
    \item $E_d$ is a set of directed edges.
    \item $E_u$ is a set of undirected edges.
    \item $E$ is a set of edges where $E = E_d \cup E_u$.
    \item $S$ is a set of strings.
    \item $V$ is a set of values of arbitrary data types.
    \item $P$ is a set of properties. Each property has the form $p = \langle k,v \rangle$, where $k \in S$ and $v \in 2^V$.
    \item $e: E \to \langle N \times N \rangle$ is a function which yields the endpoints of each directed or undirected edge (if the edge is directed, the first node is the source and the second node is the destination).
    \item $l_n : N \to 2^S$ is a function mapping each node to its multiple labels.
    \item $l_e : E \to 2^S$ is a function mapping each edge to its multiple labels.
    \item $p_n : N \to 2^P$ is a function used to assign nodes to their multiple properties.
    \item $p_e : E \to 2^P$ is a function used to assign edges to their multiple properties.
\end{enumerate}
\end{defi}

\section{Serialization}
According to our definition of the property graph model, we propose serialization in flat text and JSON. The flat text format (PG) is better for human readability and line-oriented processing, while the JSON format (JSON-PG) is best used for server-client communication.

The flat text PG format has the following characteristics, and an example is given in Figure~\ref{fig:example-pg}.

\begin{itemize}
    \item Each line describes a node or an edge.
    \item All elements in each line are separated by space or tab.
    \item In each of the node lines, the first column contains the node ID.
    \item In each of the edge lines, the first three columns contain the source node ID, the direction, and the destination node ID.
    \item Each line can contain an arbitrary number of labels.
    \item Each line can contain an arbitrary number of properties (key-value pairs).
\end{itemize}

More formally, we describe the PG format in the EBNF notation as follows.

\begin{defi}[PG Format]
\leavevmode \vspace{1mm} \\
\emph{EBNF notation of the PG format.}
\begin{scriptsize}
\begin{verbatim}
    PG         ::= (Node | Edge)+
    Node       ::= NODE_ID Labels Properties NEWLINE
    Edge       ::= NODE_ID Direction NODE_ID Labels Properties NEWLINE
    Labels     ::= Label*
    Properties ::= Properties*
    Label      ::= ':' STRING
    Property   ::= STRING ':' Value
    Value      ::= STRING | NUMBER
    Direction  ::= '--' | '->'
\end{verbatim}
\end{scriptsize}
\end{defi}

Next, we describe the JSON-PG format which follows the JSON syntax in addition to the above definition of the property graph model. The JSON-PG format has the following characteristics, and an example of the format is shown in Figure~\ref{fig:example-json}. It is to be noted that, whereas the set of labels or property values are represented as arrays in JSON, those elements are supposed to have no specific order according to the the property graph model.

\begin{itemize}
    \item Nodes and edges are listed under \texttt{nodes} and \texttt{edges} elements, respectively.
    \item Edge direction is defined with the boolean element \texttt{undirected}. By default it is false (directed).
    \item Labels are listed under the \texttt{labels} element.
    \item Properties (key-value pairs) are listed under the \texttt{properties} element.
\end{itemize}

Furthermore, we have implemented command-line tools to convert between PG and JSON-PG, as well as to transform them into formats for well-known graph databases such as Neo4j, Oracle Labs PGX, and Amazon Neptune. The practical use cases of our tools demonstrate that the proposed data model and formats have the capability to describe property graph data begin used in existing graph databases (see \url{https://github.com/g2glab/pg}).

\section{Conclusion}
In this work, we redefined the property graph model independent of specific graph database implementations and also proposed serialization formats based on the data model. Further, we implemented practical tools to convert our formats into existing ones. Our model and serialization will increase the interoperability of existing graph databases and make it easier for users to create, exchange, and utilize property graph data.

\begin{figure}[!t]
\begin{scriptsize}
\begin{verbatim}
# NODES
101  :Person  name:Alice  age:15  country:"United States"
102  :Person  :Student  name:Bob  country:Japan  country:Germany

# EDGES
101 -- 102  :sameSchool  :sameClass  since:2012
102 -> 101  :likes  since:2015
\end{verbatim}
\end{scriptsize}
\caption{Example of PG}
\label{fig:example-pg}
\end{figure}

\begin{figure}[!t]
\begin{scriptsize}
\begin{verbatim}
{
  "nodes":[
    {
     "id":101,
     "labels":["Person"],
     "properties":{"name":["Alice"], "age":[15], "country":["United States"]}
    },
    {
     "id":102,
     "labels":["Person", "Student"],
     "properties":{"name":["Bob"], "country":["Japan", "Germany"]}
    }
  ],
  "edges":[
    {
     "from":101,
     "to":102,
     "undirected":true,
     "labels":["sameSchool", "sameClass"],
     "properties":{"since":[2012]}
    },
    {
     "from":102,
     "to":101,
     "labels":["likes"],
     "properties":{"since":[2015]}
    }
  ]
}
\end{verbatim}
\end{scriptsize}
\caption{Example of JSON-PG}
\label{fig:example-json}
\end{figure}

%
%
%
%

\end{document}